\documentclass[prl,aps,twocolumn,floats,nofootinbib,showpacs]{revtex4}
\usepackage{amsmath,amssymb,graphicx,psfrag}

\begin{document}
\renewcommand{\ni}{{\noindent}}
\newcommand{\dprime}{{\prime\prime}}
\newcommand{\be}{\begin{equation}}
\newcommand{\ee}{\end{equation}}
\newcommand{\bea}{\begin{eqnarray}} 
\newcommand{\eea}{\end{eqnarray}}
\newcommand{\la}{\langle}
\newcommand{\ra}{\rangle} 
\newcommand{\dg}{\dagger}
\newcommand\lbs{\left[}
\newcommand\rbs{\right]}
\newcommand\lbr{\left(}
\newcommand\rbr{\right)}
\newcommand\f{\frac}
\newcommand\e{\epsilon}
\newcommand\ua{\uparrow}
\newcommand\da{\downarrow}
\title{Nature of single particle states in disordered graphene}
\author{Sabyasachi Nag$^{1}$} \author{Arti Garg$^{1}$}\email[Corresponding Author: ]{arti.garg@saha.ac.in}\author{T. V. Ramakrishnan$^{2,3}$} 
\affiliation{$^{1}$ Condensed Matter Physics Division, Saha Institute of Nuclear Physics, 1/AF Bidhannagar, Kolkata 700 064, India \\
$^{2}$ Department of Physics, Indian Institute of Science, Bangalore 560 012, India \\ 
$^{3}$ Department of Physics, Banaras Hindu University, Varanasi 221 005, India}
\vspace{0.2cm}
\begin{abstract}
\vspace{0.3cm}
We analyze the nature of the single particle states, away from the Dirac point, in the presence of long-range charge impurities in a tight-binding model for electrons on a two-dimensional honeycomb lattice which is of direct relevance for graphene. For a disorder potential $V(\vec{r})=V_0\exp(-|\vec{r}-\vec{r}_{imp}|^2/\xi^2)$, we demonstrate that not only the Dirac state but all the single particle states remain extended for weak enough disorder. Based on our numerical calculations of inverse participation ratio, dc conductivity, diffusion coefficient and the localization length from time evolution dynamics of the wave packet, we show that the threshold $V_{th}$ required to localize a single particle state of energy $E(\vec{k})$ is minimum for the states near the band edge and is maximum for states near the band center, implying a mobility edge starting from the band edge for weak disorder and moving towards the band center as the disorder strength increases. This can be explained in terms of the low energy Hamiltonian at any point $\vec{k}$ which has the same nature as that at the Dirac point. From the nature of the eigenfunctions it follows that a weak long range impurity will cause weak anti localization effects, which can be suppressed, giving localization if the strength of impurities is sufficiently large to cause inter-valley scattering. The inter valley spacing $2|\vec{k}|$ increases as one moves in from the band edge towards the band center, which is reflected in the behavior of $V_{th}$ and the mobility edge.

\vspace{0.cm}
\end{abstract} 
\pacs{72.80.VP,73.20.Fz, 72.15.Rn, 73.22.Pr}
\maketitle
\section{Introduction}
Graphene provides a two-dimensional (2D) electron system which is different from conventional 2D systems in many ways. Graphene can be described by a tight- binding model on a 2D honeycomb lattice~\cite{Wallace} and the energy dispersion is linear in momentum near the Fermi points at half-filling resulting in Dirac like quasi-particles~\cite{Castro_neto,Peres}. 
It is well known that the Dirac states in graphene evade Anderson localization in the presence of weak long range charge impurities~\cite{Bardarson,Mirlin,SDasSarma}. Only short range impurities or very strong long range impurities, either of which can cause scattering from one Dirac valley to another, show weak localization correction to conductivity rather than weak-anti localization~\cite{Surat-Ando}. But the nature of single particle states away from the Dirac point has not been explored in detail so far. Does a higher energy state, away from the Dirac point, get localized in the presence of an infinitesimal strength of disorder, as expected for a conventional 2D system from the scaling theory of localization~\cite{scaling_th} or does it evade localization like a Dirac state? In this paper we focus on these questions in detail.

An impurity which is long range in real space can not induce large momentum transfer in a scattering event. Therefore, in the presence of weak long range impurities, only intra valley scattering around a Dirac point is possible. A backscattering event around a Dirac point is further suppressed due to Berry phase effect of Dirac states~\cite{Surat-Ando}. 
A near backscattering event, then gives a positive correction to the conductivity, that is, $\Delta \sigma = +ln{L/l_e}$ resulting in a postitive $\beta$ function and weak-anti localization effect. Thus, although in a conventional 2-dimensional system, all single particle states are localized in the presence of an infinitesimal disorder~\cite{scaling_th}, Dirac states in graphene stay extended for weak long range impurities. 
Though ``trigonal warping" of the spectrum, obtained by keeping terms up to second order in momentum, is necessary to understand the physics of WAL in this model fully~\cite{Altshuler} and can suppress WAL effects under certain conditions, for graphene WAL has been observed experimentally~\cite{WAL} and numerically~\cite{Bardarson,Beenakker,SDasSarma,Castro,Roche2}.
 When the long range impurity potential, increases in strength, such that the scattering amplitude for intervalley scattering becomes comparable to the intra valley scattering amplitude (as shown in~\cite{zhang}), transition from WAL to weak localization takes place~\cite{Bardarson,Roche2}. 

In this work, we study the nature of all the single particle states of the tight binding model on the honeycomb lattice in the presence of long range charge impurities. For the disorder potential ($V(\vec{r})=V_0 \exp(-|\vec{r}-\vec{r}_{imp}|^2/\xi^2$, which is a widely used model of impurities for graphene ~\cite{Bardarson,SDasSarma,zhang,Castro,Beenakker,Lewenkopf}, we demonstrate that all single particle states remain extended for weak enough strength of disorder. We show clear evidence in support of this from analysis of inverse participation ratio (IPR) and the localization length obtained from the time evolution dynamics of the wave packet.
We show that the conductivity increases with the disorder strength and the system size in the weak disorder regime, not only for the charge neutral graphene but also for highly doped graphene. 
There is a threshold disorder $V_{th}$ beyond which an energy state shows localized trend. $V_{th}$ is small for states near the band edge and is maximum for states near the band center which implies the appearance of a mobility edge (ME) in this system. We also provide an intuitive explanation for this surprising result from an effective Hamiltonian calculation. We derive the low energy Hamiltonian for any $\vec{k}$ point on the Brillouin zone and show that it has the same nature as the low energy Hamiltonian at the Dirac point. From the nature of the Hamiltonian and the eigenfunctions, it follows that (as shown e.g. by Ando~\cite{Surat-Ando} even before the discovery of graphene) a random potential will cause weak anti localization effects; these can be suppressed, giving localization~\cite{Surat-Ando} if the impurities can cause sufficient inter-valley ($\vec{k} \Rightarrow -\vec{k}$) scattering.  Since the inter valley spacing $2|\vec{k}|$ increases as one moves in from the band edge towards the band center, $V_{th}$ required to localize a single particle state also increases implying the formation of the mobility edge starting from the band edge and moving in, with increasing overall strength of the long range potential.  

The rest of the paper is organized as follows. In section I, we describe the model and the method used to study followed by the details of numerical results on IPR, conductivity and diffusion coefficient in section II. In section III, we describe the effective Hamiltonian picture away from the Dirac point in support of our numerical results and end this paper with conclusions and discussions.  

\section{Model and Method}
The disorder model we study is
\be
H=-t_0\sum_{\langle ij \rangle ,\sigma}[a^\dagger_{i,\sigma}b_{j,\sigma} + h.c.] +\sum_i V(i) n(i) -\mu \sum_i n(i)
\label{hamil}
\ee
Here $t_0$ is the nearest neighbor hopping amplitude, $\mu$ is the chemical potential which can be tuned to fix the average particle density in the system. At half filling $\mu=0$. 
$V(i)$ is the random on site potential due to smooth scatterers considered. 
$V(i)$ is the sum of contributions from $n_{imp} \times N$ impurities located randomly at $\vec{r}_{imp}$ among $N=2L^2$ lattice sites, with $L$ being the length of one sub lattice, so that $V(i)=\sum_{\vec{r}_{imp}}V_0(\vec{r}_{imp})\exp[-|\vec{r}_i-\vec{r}_{imp}|^2/\xi^2]$, where $V_0(\vec{r}_{imp})$ is randomly distributed within $[-V,V]$.  We exactly diagonalize the $ N \times N$ Hamiltonian matrix for a given disorder configuration. Once we have all the $N$ eigenvalues, we can calculate physical quantities of interest exactly for that disorder configuration. Data is obtained for various independent disorder configurations and all physical quantities are obtained by doing averaging over many independent disorder configurations. We mainly focus on inverse participation ratio and dc conductivity which is calculated using the Kubo formula. We also study time evolution dynamics of the wave packet to analyze the nature of single particle states.

\section{Results}

The results described in the following sections are for $n_{imp}=10\%$ and the range of the impurity potential $\xi=3a$ where $a$ is the inter atomic distance.  We first present results for inverse participation ratio which is crucial to analyze whether a single particle state is extended or localized. In the following subsections we present results for the dc conductivity and time evolution dynamics study respectively. 

\begin{figure}[h!]
\begin{center}
\includegraphics[width=2.5in,angle=-90]{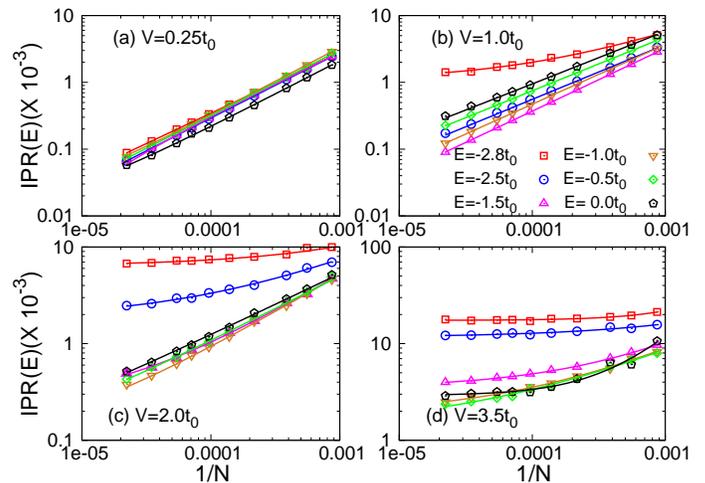}
\caption{[a]:For $V=0.25t_0$, $IPR(E) \sim L^{-d^\star(E)}$ for all $E$ states which implies that these states are extended. 
As $V$ increases, first states near the band edge show $IPR(E)\sim IPR_0+\alpha(E)L^{-d^\star(E)}$ as shown in Fig[b] for $V=1.0t_0$. With increasing $V$ further, more states $(E\in[-2.8:1.5]t_0)$ in the middle of the band gets localized as shown in [c] for $V=2.0t_0$.  Finally all the states get localized, including the zero energy state as shown in [d].
This data has been obtained by doing averaging over 50-100 independent disorder configurations and the energy bin used is $dE \in [0.03,0.075]$ for various $E$ values.}
\label{IPR}
\vskip-1cm
\end{center}
\end{figure}

\subsection{Inverse participation ratio (IPR)} 
 To decide whether a single particle state is localized or not, we calculate the IPR which is defined as
\be
IPR(E) = \langle \sum_{i,n} |\Psi_n(i)|^4 \delta(E-E_n) \rangle _C
\ee
where $\Psi_n(i)$ is an eigenfunction (normalized) with eigenvalue $E_n$ of the Hamiltonian in Eq.~\ref{hamil} and $\langle \rangle _C$ indicates the configuration averaging.  In our numerical calculation of IPR(E), we replace the delta function by a box distribution of finite width $d E$. IPR of a state is inversely proportional to the portion of space which ``participates'' in that eigenstate.  For plane waves $IPR(E) \sim L^{-d}$ for a d dimensional system of length $L$.  But for a general extended state $IPR(E) \sim L^{-d^\star(E)}$ where $d^\star(E) \le d$~\cite{Kramer}. Therefore, in the thermodynamic limit, IPR vanishes for an extended state while it is non-zero, having the form $IPR(E) = IPR_0(E)+\alpha(E) L^{-d^\star(E)}$, for a localized state. 
Fig.~\ref{IPR} shows $IPR(E)$ for various energy states as a function of $1/N$. At $V=0.25t_0$, $IPR(E)$ for all the states goes as $L^{-d^\star(E)}$ implying that all these states are extended (for details, see Appendix A). As $V$ increases, first states near the band edge get localized and then the states in the middle of the band get localized.  Eventually at $V_{th}=2.5t_0$, IPR in the thermodynamic limit becomes non zero for all the states. Our scaling analysis, which shows that the Dirac state ($E=0$) is extended below $V=2.5t_0$ and is localized for larger $V$ values, is consistent with earlier work on magneto conductance calculation where WAL to WL transition was observed for charge neutral graphene at $V=2t_0$~\cite{Roche2}. This provides a check for our scaling analysis.  

\begin{figure}[h!]
\begin{center}
\includegraphics[width=1.75in,angle=-90]{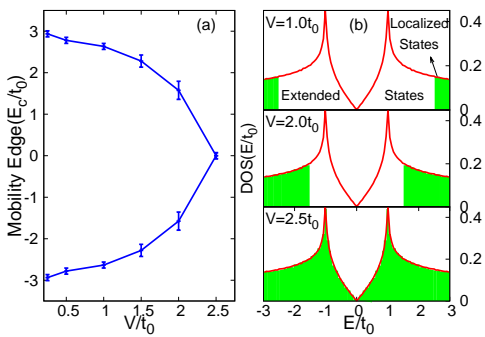}
\caption{The mobility edge $E_c(V)$ vs $V$, obtained from the IPR analysis. The mobility edge starts from the band edge ($E=\pm 3t_0$) for weak disorder and moves towards the band center ($E=0$) as $V$ increases. Right panel shows the density of states for clean case to schematically show the location of the ME.}
\label{ME}
\vskip-1.8cm
\end{center}
\end{figure}

The IPR analysis provides a clear evidence that not only the Dirac states but all the single particle states remain extended for very weak long range disorder. A threshold disorder $V_{th}(E)$ is required to localize a state at energy $E$. $V_{th}(E)$ is larger for a state close to the band center as compared to its value for a state near the band edge. Fig.~\ref{ME} shows how the mobility edge $E_c$, which separates localized states from extended states, moves as $V$ increases. The results shown are for quite big system sizes (maximum $N = 2 \times 150^2$) and disorder averaging is done over 50-100 configurations. Density plots of the wave functions also support the IPR analysis. 

\begin{figure}
\begin{center}
\includegraphics[width=2.25in,angle=-90]{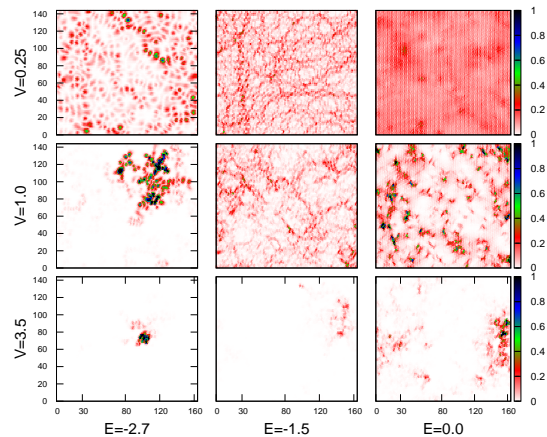}
\caption{Density plots of $|\Psi_n(i)|^2$, for $L=96$ for a single disorder configuration, where $\Psi_n(i)$ is the wave function for the eigenstate with energy $E_n$. For $V=0.25t_0$ all states are extended over the entire lattice. For $V=1.0t_0$, state near the band edge $E_n =-2.7t_0$ has little spatial extension while the lower energy states are extended over the entire lattice. Dirac state ($E=0$) gets localized at a much larger value of $V$ as shown in the bottom panel.}
\label{wf}
\end{center}
\end{figure}

Right panel of Fig.~\ref{wf} shows the spatial extent of the wave functions $\Psi_n(i)$ for an eigenstate with energy $E_n$ using the density plots of $|\Psi_n(i)|^2$ for a specific disorder configuration. One can see that at weak disorder, e.g., $V=0.25t_0$ for all the states, including those near the band edge (e.g., $E_n = -2.7t_0$), $|\Psi_n(i)|^2$ is extended over the entire lattice. Further, states close to the band edge show fractal nature even at $V=0.25t_0$ while states near the band center are very uniformly distributed over the entire lattice, which is consistent with the values of $d^\star$ obtained from IPR analysis.
For $V=1.0t_0$,  $|\Psi_n(i)|^2$ for $E_n\sim 2.7t_0$ has weight only at a few points on the lattice as shown in Fig.~\ref{wf}. Also notice that fractal nature of the wave functions becomes dominant for all the energy states as the disorder strength $V$ increases. Energy state at $E=0$ gets localized at comparatively larger disorder value as shown in the bottom panel. This is consistent with our IPR analysis. Left panel of Fig.~\ref{ME} shows the schematic plot of the mobility edge obtained from the IPR analysis.

\subsection{Conductivity}
In this section we present results for the regular part of the dc conductivity $\sigma$ calculated using the Kubo formula details of which are given in Appendix B. Conventionally, in a metallic system scattering with impurities reduces the conductivity. On the contrary our calculation shows an increase in $\sigma$ with disorder in the limit of weak disorder. 
\begin{figure}[!htb]
\begin{center}
\vskip-0.2cm
\includegraphics[width=2.5in,angle=-90]{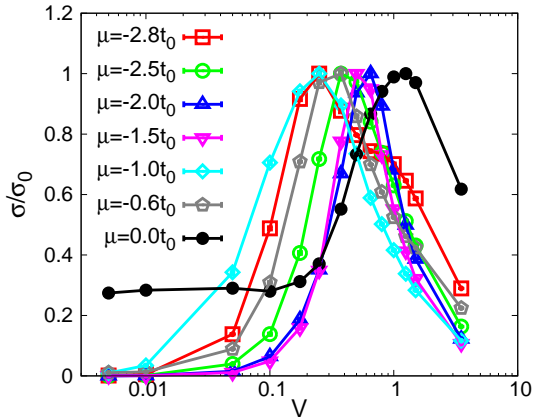}
\caption{ $\sigma/\sigma_{0}$ vs $V$ for various values of $\mu$ where $\sigma_0$ is the maximum value of $\sigma$ as a function of $V$ for the corresponding $\mu$. For all values of $\mu$, $\sigma$ first increases with $V$ followed up by its decrease as $V$ increases further. 
Data shown here has been obtained by doing averaging over 60 independent disorder configurations.}
\label{sigma1}
\vskip-0.4cm
\end{center}
\end{figure}
As shown in Fig.~\ref{sigma1}, $\sigma$ first increases with $V$, reaches a maximum followed up by its decrease on further increase in $V$.  Increase in $\sigma$ with $V$ has been observed for graphene at the charge neutral point $\mu=0$, where the Fermi surface consists of Dirac points ~\cite{Bardarson,Beenakker,Castro,SDasSarma}, and is explained in terms of impurity assisted tunneling~\cite{Titov}. This relies upon the symplectic symmetry of the Dirac Hamiltonian~\cite{Titov} which is preserved in the presence of weak long range impurities due to the absence of inter-valley scattering. Our results show that even when the system is doped far away from the charge neutral point, $\sigma$ shows a trend similar to that of the charge neutral graphene. As we will show in section III, this happens because the effective Hamiltonian away from the Dirac point still belongs to the symplectic symmetry class in the presence of weak long range impurities.
 
\begin{figure}[!htb]
\begin{center}
\includegraphics[width=2.0in,angle=-90]{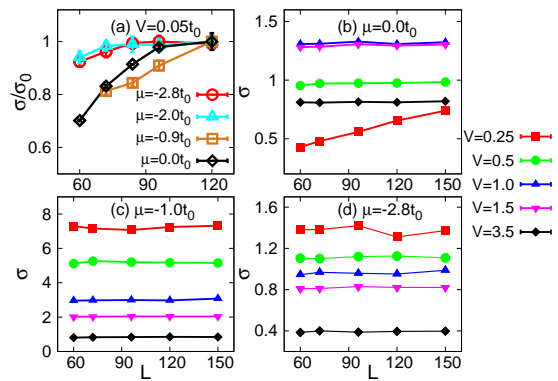}
\caption{Panel [a]: $\sigma/\sigma_{0}$ vs $L$ for $V=0.05t_0$ where $\sigma_0$ is the maximum value of $\sigma$ as a function of $L$. In the weak disorder limit, $\sigma$ increases with $L$ not only for $\mu=0$ but also for $\mu$ values far away from the charge neutral point. Panel [b-d]: $\sigma$ in units of $\frac{e^2}{h}$ as a function of $L$. For $\mu=0$, $\sigma$ increases with $L$ even for intermediate disorder strength $V \le 0.5t_0$).  For all other $\mu$ values and for other $V$ values, except for some small zigzag oscillation $\sigma$ is effectively constant. These results are obtained by doing averaging over 60-100 disorder configurations from large to small $L$.}
\label{sigma2}
\vskip-0.2cm
\end{center}
\end{figure}

Panel [a] in Fig.~\ref{sigma2} shows $\sigma$ vs the system size $L$ for $V=0.05t_0$. In this extreme weak disorder limit, $\sigma$ increases with the system size, not only for $\mu=0$, which has been seen in earlier work~\cite{Bardarson,Castro,SDasSarma,Surat-Ando}, but also for highly doped systems which demonstrates that higher energy states are similar in nature to the Dirac states. Like Dirac states, higher energy states also remain extended for weak disorder and show weak anti localization effects.    

Upon increasing $V$ further, for small $\mu$ values, $\sigma$ continues to show an increase with $L$ up to intermediate disorder strength ($V\sim 0.5t_0$), implying a positive $\beta$ function, (see Fig.~\ref{sigma2}[b]) which is consistent with conclusions from earlier work~\cite{Surat-Ando,Bardarson,Lewenkopf,SDasSarma}. We see signatures of positive $\beta$ function for $\mu=0.25t_0$ as well for $V=0.25t_0$. But for higher values of $\mu$, for intermediate disorder $\sigma$ remains constant w.r.t $L$. This is because $V_{th}(E)$ for higher $E$ states is smaller than that for the lower $E$ states, and $V=0.25t_0$ is not weak compared to $V_{th}(E)$ for these states. 
For higher disorder strength, for all $\mu$ values, $\sigma$ is more or less constant for the system sizes studied as shown in Fig.~\ref{sigma2}. But eventually for large disorder $\sigma$ should decrease with $L$ as the system enters in the localized regime which could not be captured by the Kubo formula though it is very clearly visible in the IPR analysis shown earlier and also in the diffusion coefficient study which we will discuss in the next section.  

\subsection{Time Evolution Dynamics}
 In this section we study the time evolution dynamics of a wave packet to analyze the effect of disorder on it~\cite{Markos,Roche}. We start with an initial wave packet $\Psi(i,t=0)$, which is an eigenstate of the Hamiltonian in Eq.~\ref{hamil} for a given value of $V$ and a particular disorder configuration on a small lattice for which the sub-lattice length is $L_{in}$. Here $i$ is the site index and $t$ is the time (in units of $1/t_0$). We place this wave packet in the middle of a bigger lattice of sub lattice length $L$, which corresponds to certain expectation value $E$ of the Hamiltonian, and evolve it in time with respect to the Hamiltonian of Eq.~\ref{hamil} for bigger system size.  
Then the wave function at any subsequent time is given by $\Psi(i, t)=\sum_n c_n \psi_n(i)e^{-iE_nt}$ where $E_n$ is an eigenvalue of the Hamiltonian for $L=150$ system and $\Psi_n(i)$ is the corresponding eigenvector.  Note that the energy of the evolving wave packet is determined as the expectation value of the Hamiltonian operator $<\Psi|H|\Psi>=E=\sum_n |c_n|^2E_n$. 

\begin{figure}[!htb]
\begin{center}
\includegraphics[width=2.5in,angle=-90]{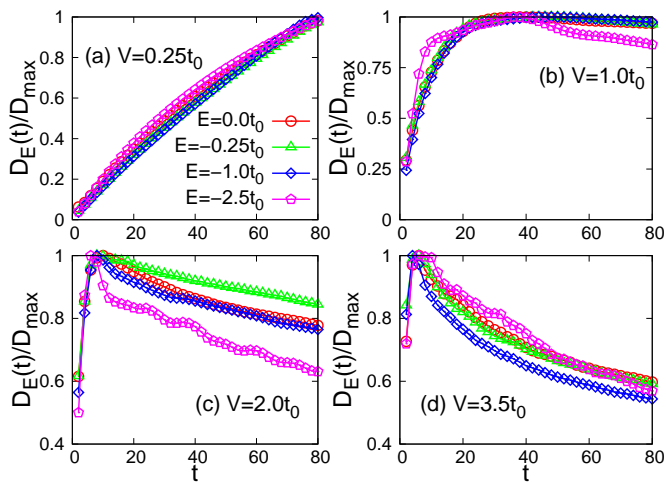}
\caption{$D_E(t)$, divided by its maximum value, vs $t$ (in units of $1/t_0$) for various values of $E$. Panel [a]: For $V=0.25t_0$, $D_E(t)$ increases almost linearly for small time and then with $t^c$ with $c<1$ for all $E$ values. Panel [b]: For intermediate disorder, for low energy states $D_E(t)$ shows saturation in the large time regime, but for higher $E$ states, $D_E(t)$ shows decreasing trend with $t$ in the long time regime. For large $V$, shown in panel [c] and [d], for all the states, $D_E(t)$ decreases with $t$ in the long time regime. Here time $t$ is in units of $\hbar/t_0$. } 
\label{diffusion}
\vskip-1cm
\end{center}
\end{figure}

We calculate the expectation value of $\la {\mathbf{r}}^2\ra$ where $\mathbf{r}$ is the position operator;
\begin{equation}
<\mathbf{r}^2(t)>_E=\sum_i \mathbf{r_i}^2|\Psi(i, t)|^2
\end{equation}
where $\mathbf{r_i}$ denotes the position of the $i-$th lattice point measured from the center of the lattice. For the clean case, where all the states are extended, $\la \mathbf{r}^2(t)\ra_E \sim t^2$. For $\la \mathbf{r}^2(t)\ra_E \propto t$, the state is diffusive with the constant of proportionality being the diffusion coefficient. For a localized state, the wave packet spreads up to the localization length and then remains constant in time $\la \mathbf{r}^2(t) \ra_E \sim t^0$ in the limit of large time. 
\\
\\
{\bf{Diffusion Coefficient}}: We define a generalized diffusion coefficient $D_E(t)$ following~\cite{diff} as  
\begin{equation}
D_E(t)=\frac{\la \mathbf{r}^2(t) \ra_E - \la \mathbf{r}^2(0)\ra_E}{t}
\end{equation}
which depends upon time $t$. We calculated $D_E(t)$ for various values of $E$ and the disorder strength $V$. Data shown in Fig.~\ref{diffusion} has been averaged over 15 independent disorder configurations. 
As shown in Fig.~\ref{diffusion}, for $V=0.25t_0$, $D_E(t)$ for all $E$ values keeps increasing with $t$ indicating their extended nature in consistency with our earlier results. Note that for $E=-2.5t_0$, $D_E(t)$ first increases with $t$ followed by a saturation of $D(E,t)$ at a maximum value $D_{max}(E)$ which indicates the diffusive behavior. For intermediate values of $V$, as shown in panel [b] for $V=1.0t_0$, $D_E(t)$, in the large time regime, shows diffusive trend for lower energy states while it decreases with $t$ for $E=-2.5t_0$ (close to the band edge) due to localization effects. The time after which $D_E(t)$ starts showing localization trend decreases with increase in disorder strength. Eventually for very high disorder values, as shown in panels [c] and [d], $D_E(t)$ decreases with $t$ in most of the time regime for all the energy states. 
\begin{figure}[!htb]
\begin{center}
\includegraphics[width=3.0in,angle=0]{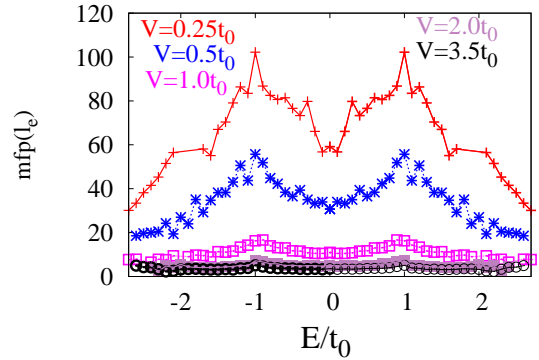}
\caption{ Elastic mean free path $l_e$ vs $E$ for various disorder strengths obtained from $D_{max}(E)$. $l_e$ decrease with increase in disorder strength. Note that for $V \ge 0.5t_0$, $l_e$ is much less than  $L$ for all values of $E$.}
\label{mfp}
\vskip-1cm
\end{center}
\end{figure}

{\bf{Elastic Mean Free path}}: For the diffusive regime, where $D_E(t)$ saturates to a maximum value $D_{max}(E)$, one can estimate the elastic mean free path $l_e$ using the relation $D_{max}(E) = v_f l_e$, where $v_f$ is the Fermi velocity. We extend this definition following ~\cite{Roche2} and calculate $l_e$ using the maximum value $D_{max}(E)$ for all disorder regimes. Fig.~\ref{mfp} shows $l_e$ vs $E$ for various values of $V$. The data shown is for $L=150$ and $L_{in}=6$. As expected, $l_e$ decreases with increase in the disorder strength. For weaker disorder values $V=0.25t_0$ and $0.5t_0$, $l_e$ shows a dip for $E=0$ while the energy dependence gets smoothened out for stronger disorder values. This energy dependence of $l_e$ for weak disorder is consistent with what has been observed earlier ~\cite{Roche2} for long range disorder.

 Most important point to notice is that even for weak disorder, $l_e \ll L$ for states near the band edge. This ensures that our analysis of IPR, which suggests that for weak disorder not only the Dirac state but also the states near the band edge remain extended, is reliable. For $V \ge 0.5t_0$, $l_e$ for all values of $E$, is smaller than all the system sizes studied for IPR analysis.    
\begin{figure}[!htb]
\begin{center}
\includegraphics[width=3.0in,angle=0]{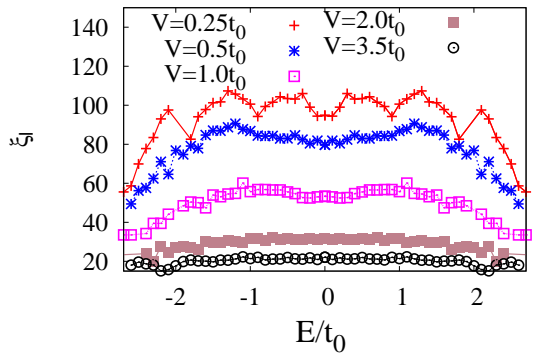}
\caption{ Localization length $\xi_l$ (in unit of a) vs $E$ for various disorder strengths. For $V=0.25t_0$ $\xi_l$ is of the order of $L$ for all values of $E$. With increase in the disorder strength $V$, $\xi_l$ decreases. For $V=3.5t_0$, $\xi_l \ll L$ for all the energy states indicating there localized nature.}
\label{xi}
\vskip-1cm
\end{center}
\end{figure}
\\
\\
{\bf{Localization Length}}: We also calculated the localization length $\xi_l$ from time evolution dynamics of the wave packet. Since $\la r^2(E,t)\ra$ gives the spread of the wave packet of energy $E$ at time $t$, maximum value of $\la r^2\ra$ provides a good estimate of $\xi_l^2$.  Fig.~\ref{xi} shows $\xi_l$ vs $E$ for various disorder values. This data is obtained for $L=150$. For $V=0.25t_0$, $\xi_l$ is of the order of the system size for all $E$ states. As the disorder strength increases, $\xi_l$ decreases eventually becoming much smaller than the system sizes studied indicating the localization behavior. 
\begin{figure}[!htb]
\begin{center}
\includegraphics[width=2.5in,angle=-90]{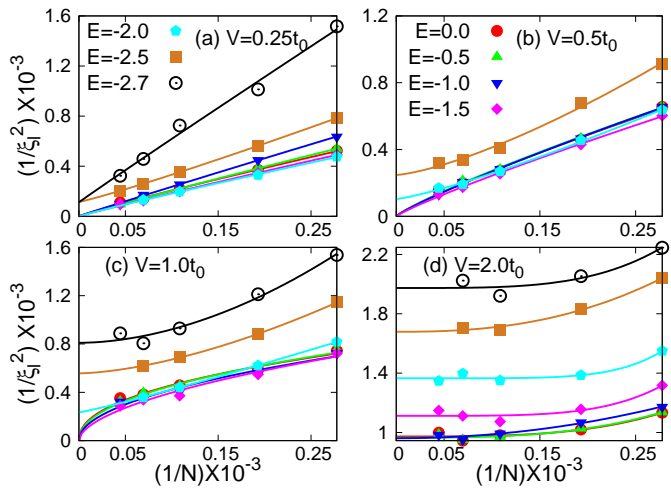}
\caption{$1/\xi_l^2$ vs $1/\L^2$ for various energy states. Panel [a] shows the results for $V=0.25t_0$ where $1/\xi_l^2 \rightarrow 0$ in the large $L$ for all $E$ states except for $E=-2.7t_0$ and $-2.5t_0$ implying that only these two states are localized at $V=0.25t_0$. For $V=0.5t_0$, $1/\xi_l^2$ remains finite in the thermodynamic limit for $E=-2.0t_0$ as well. Eventually for very large disorder, for all $E$ states $1/\xi_l^2$ remains finite in the thermodynamic limit.}
\label{xi2}
\vskip-1cm
\end{center}
\end{figure}

To ensure whether an energy state is extended for weak enough disorder or not, we did system size analysis of $\xi_l$. In Fig.~\ref{xi2}, we have plotted $1/\xi_l^2$ vs $1/\L^2$, in analogy with IPR plots. Since $\xi_l^2$ gives the area enclosed by a wave packet, $1/\xi_l^2$ is proportional to IPR. As seen in Fig.~\ref{xi2}, for $V=0.25t_0$, $\frac{1}{\xi_l^2}\rightarrow 0$ in the thermodynamic limit for all $E$ states except for $E=-2.7t_0$ and $-2.5t_0$. Let $\xi_0$ is the extrapolated value of $\xi_l$ in the thermodynamic limit.  Hence, for $E=-2.7$ and $-2.5t_0$, $\xi_0$ is finite indicating the localized nature of these states while for all other energy states $\xi_0$ is infinite indicating their extended nature. As $V$ increases, more energy states in the band show localized behavior. For example for $V=0.5t_0$, even $E=-2.0$ shows finite $\xi_0$. Eventually for very large disorder $V=2.0t_0$, $\xi_0$ is finite for all the energy states. 
\begin{figure}[!htb]
\begin{center}
\includegraphics[width=2.75in,angle=0]{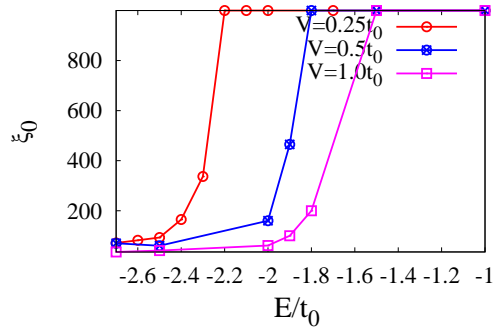}
\caption{$\xi_0 = lim_{L\rightarrow \infty}\xi_l$ (in units of a) vs $E$ for various disorder strengths. For $V=0.25t_0$, $\xi_0$ is finite for $ |E|> 2.3t_0$ while it is infinite for $|E| < 2.2t_0$. For higher disorder values, $\xi_0$ remains finite for a broader range of energy states near the band edge, e.g., for $V=1.0t_0$, $\xi_0$ is finite for $|E| > 1.8t_0$. Wherever $\xi_0$ is infinity, it has been replaced by 1000a for the sake of presentation.}
\label{xi3}
\vskip-1cm
\end{center}
\end{figure}

Fig.~\ref{xi3} shows $\xi_0$ vs $E$ for various values of disorder. Here for the sake of presentation, wherever $\xi_0$ is infinity, we represent it by 1000a. As one can see, for $V=0.25t_0$, $\xi_0$ diverges after $E=-2.5t_0$. For $V=0.5t_0$, $\xi_0$ is finite for a longer range of energy inside the band and shows divergence only for $E>-2.0t_0$. This analysis supports our proposal and IPR analysis that all energy states remain extended below certain threshold disorder strength resulting in the appearance of a mobility edge in this system~\cite{footnote}. 
\section{Effective Hamiltonian}
In this section we provide an explanation for our numerical findings by looking at the effective low energy Hamiltonian. 
The tight binding Hamiltonian on the honeycomb lattice can be written as 
\be
H=\sum_{\vec{k},\sigma}\Phi(\vec{k}) a^\dagger_{\sigma}(\vec{k})b_\sigma(\vec{k})+h.c.
\label{h0}
\ee 
with $\Phi(\vec{k})=-t_0\sum_{i=1,3} \exp(i \vec{k}\cdot\vec{\delta_i})$. Since the spin symmetry is preserved in the presence of impurities, we will ignore real spin index $\sigma$ from now onwards. Here $\vec{\delta_i}$ are the nearest neighbor lattice vectors. In the vicinity of a point $(k_x,k_y)$ on an equal energy contour of energy $E$, one can do the Taylor series expansion to get $\Phi(k_x+q_x,k_y+q_y) \sim (a-ib)+(\gamma_1 q_x+\gamma_2 q_y) -i(u_1 q_x+u_2q_y)$ for small $q_x,q_y$ where $a,b,\gamma_{1,2}$ and $u_{1,2}$ are functions of $(k_x,k_y)$. Thus the low energy Hamiltonian around a point on an equal energy contour corresponding to $E\sim \pm\sqrt{a^2+b^2}$ is given by 
\be
H_{eff}=H-H_0\sim (\gamma_1 q_x+\gamma_2 q_y)\sigma_x + (u_1 q_x+u_2 q_y)\sigma_y
\label{Heff1}
\ee
where $\sigma_{x,y}$ are the Pauli matrices and $H_0=a\sigma_x+b\sigma_y$.  
Note that like the Dirac Hamiltonian, this Hamiltonian is invariant under pseudo-spin reversal symmetry $\sigma_y H^T_{eff}\sigma_y$~\cite{Mirlin,SDasSarma}.
Around the diagonally opposite point $(-k_x,-k_y)$ on this energy contour $E$, the coefficients $\gamma_{1,2}$ change sign while $u_{1,2}$ remain same. 
Eigenfunctions for the Hamiltonian are given by
\bea
\Psi^{\vec{k}}_{\pm}(\vec{q})=\f{\exp( i\theta(\vec{q})/2)}{\sqrt{2}}\lbr \begin{array} {c} \exp( i\theta(\vec{q})/2) \\ \pm \exp(-i\theta(\vec{q})/2) \end{array}\rbr\\
\label{eigenfn}
\eea
where $\pm$ correspond to positive and negative eigenvalues and $\theta\sim\theta_0+\delta\theta$ with $tan(\delta\theta(\vec{q}))=\frac{u_1q_x+u_2 q_y}{\gamma_1 q_x+\gamma_2 q_y}$ and $tan(\theta_0) = \frac{b}{a}$. Just like for the Dirac states, due to the non trivial Berry phase effect of the wavefunction, a back scattering process around the point $\vec{k}$ and its time reversed process are out of phase with each other causing suppression of the back scattering process. A near back scattering event, again due to the Berry phase, then gives rise to the opposite sign of the conductivity correction~\cite{Surat-Ando} resulting in WAL effect. 
Note that since we have kept only first order terms in the Taylor expansion, this picture will hold only in close vicinity of the point $\vec{k}$. Thus only for very weak disorder, which can induce small momentum transfer scatterings, this picture holds true. But it conveys the main message that for extremely weak disorder, all states will show WAL effect. 

\begin{figure}
\begin{center}
\includegraphics[width=1.8in,angle=-90]{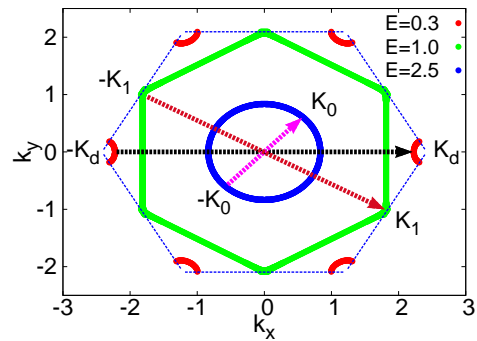}
\caption{Equal energy contours for various  values of $E/t_0$ for the clean case. The momentum transfer required to go from a point $\vec{k}$ on an energy contour of energy $E$ to $-\vec{k}$, is smaller for $E$ close to the band edge ($E=\pm 3t_0$) and increases as $E$ moves towards the band center ($E=0$) for which energy contour shrinks to six Dirac points.} 
\label{contours}
\vskip-0.2cm
\end{center}
\end{figure}

 A strong long range impurity which can result in momentum transfer $2\vec k$ can scatter an electron from point $\vec{k}$ on the energy contour to the diagonally opposite point $-\vec{k}$, and give rise to weak localization effects. The strength of disorder required to cause scattering from $\vec{k}$ to $-\vec{k}$ depends on the energy contour. For energies close to the band edge ($E\sim \pm 3t_0$), where the energy contours have very small radius (Fig.~\ref{contours}), small disorder is sufficient to cause this scattering and this value of disorder strength increases as one moves away from the band edge towards the band center ($E\sim 0$). This intuitive picture supports the results obtained from our numerical calculations. 
\section{Conclusions and Discussions}
 In conclusion, we have analyzed the nature of single particle states in the tight binding honeycomb lattice in the presence of long range charge impurities and demonstrated that not only the Dirac state but all the single particle states remain extended in the presence of weak long range impurities due to generic anti localization effects and prohibited backscattering events. This is because the nature of low energy Hamiltonian and the wave functions are basically of the same nature as for the Dirac states. 
A threshold strength $V_{th}$ of disorder is required to localize a state with energy $E$ since the random  potential needs to scatter an electron from point $\vec {k}$ on an equal energy contour of energy $E$ to point $-\vec{k}$ on it. Therefore, though states near the band center remain extended for weak to intermediate values of disorder strength, states near the band edge show extended nature only for weak disorder. 
We do see clear indications of this in the analysis of inverse participation ratio (IPR), charge conductivity, the generalized diffusion coefficient $D_E(t)$ and the localization length obtained from time evolution dynamics of the wave packet. 

In this work we presented all the results for a specific choice of the range of impurity potential $\xi$ and the concentration of impurities $n_{imp}$. Location of the mobility edge for a given value of $V$ also depends upon $\xi$ and $n_{imp}$. For larger values of $\xi$, we should expect ME to be moving slowly with the disorder strength $V_0$. In principle, since screening of the potential depends upon the density of states, effective range of the potential $\xi$ itself is a function of the doping. It will be interesting to carry out a full self-consistent calculation for a model of long range impurity potential where range of the potential is decided by the doping. These are questions for future work. 

At this end, we would like to mention that for simplicity we have worked with the nearest neighbor  tight binding (NNTB) model of graphene which is a very good model at low energy (near Dirac states). Also the 3rd NNTB model does not make much of a difference in the dispersion with respect to what we have for the NNTB model~\cite{Roche3}. Only  the exact band structure calculation shows that there are new somewhat non dispersive high and low energy bands near the gamma point~\cite{Roche3}. There is no specific reason why these new sub bands should be particularly effective for localization. 
There cannot be any decay a la Fermi Golden rule, for states at the band edges, which are our main points.   

We hope our study will motivate experiments on highly doped graphene to look for extended nature of single particle states away from the Dirac point. 
\section{Acknowledgements}
 S. N would like to acknowledge Kausik Das (SINP) for computational help. 
\section{Appendix A}
In this Appendix, we give details of IPR analysis. 
For plane waves $IPR(E) \sim L^{-d}$ for a d dimensional system of length $L$.  But for a general extended state $IPR(E) \sim L^{-d^\star(E)}$ where $d^\star(E) \le d$. Therefore, in the thermodynamic limit, IPR vanishes for an extended state while it is non-zero, having the form $IPR(E) = IPR_0(E)+\alpha(E) L^{-d^\star(E)}$, for a localized state. Below we tabulate the values of $d^\star(E)/2$ for various values of energy states $E$ and the disorder strength $V$.
\begin{center}
 \begin{tabular}{||c c c c c c c ||}
 \hline
 V & E=2.8 & E=2.5 & E=1.5 & E=1.0 & E=0.5 & E=0.0 \\ [0.5ex]
 \hline
 0.25 & 0.91 & 0.98 & 0.95 & 1.0 & 0.98 & 0.99 \\
 \hline
 0.5 & 0.81 & 0.95 & 0.95 & 0.94 & 0.96 & 0.76 \\
 \hline
 1.0 & 0.69 & 0.82 & 0.94 & 0.89 & 0.80 & 0.79 \\
 \hline
 1.5 & 0.43 & 0.69 & 0.79 & 0.81 & 0.80 & 0.75 \\
 \hline
 2.0 & 0.60 & 0.57 & 0.75 & 0.78 & 0.69 & 0.64 \\
 \hline
 2.5 & 0.14 & 0.64 & 0.67 & 0.67 & 0.72 & 0.72 \\ [1ex]
 \hline
\end{tabular}
\end{center}
As shown in this table, for $V=0.25t_0$, $d^\star \sim d=2$ for states away from the band edge because $V=0.25t_0$ is weak disorder for the single particle states away from the band edge, but states near the band edge have $d^\star$ values quite off from $d=2$. As $V$ increases, for almost all the energy states, $d^\star$ decreases and becomes significantly smaller than $d=2$.
\section{Appendix B}
In this Appendix, we will provide details of calculation of conductivity within the Kubo formalism ~\cite{Mahan} which is given by
\begin{equation}
\sigma_{xx}=-lim_{\omega \rightarrow 0} lim_{q\rightarrow 0}\frac{Im[\Lambda_{xx}(q,\omega)]}{\hbar\omega 2L^2}
\label{sigma}
 \end{equation}
Here $\Lambda_{xx}(q,\omega)$ is the current-current correlation function, given by
\be
\Lambda_{xx}(q,\omega)=\int_0^{\infty} e^{i\omega t} \la [J_x(q,t),J_x(-q,0)]\ra
\ee
with 
\be
J_x=-iet_0\sum_{<i,j>,\sigma}\lbs a^\dagger_{i\sigma}b_{i+j,\sigma}-b^\dagger_{i+j,\sigma}a_{i,\sigma}\rbs\delta_{i,j}^x
\ee
 and $\delta_{i,j}^x$ being the x component of the nearest neighbor vector $\vec{\delta}$ between sites $i$ and $j$. Here  $\sigma$ is the index for electron spin and since there is a spin degeneracy in the system, each spin makes equal contribution to the charge conductivity. Thus from now onwards, we will skip the spin index.

The tight binding Hamiltonian in Eq.[\ref{hamil}] in the presence of disorder can be diagonalized using the transformation
\bea
 a_{i}=\sum\limits_{n=1}^{L^2}[\psi_n^a(i)A_{n}+\psi_{n+L^2}^a(i)B_{n}] \nonumber \\
 b_{i}=\sum\limits_{n=1}^{L^2}[\psi_n^b(i)A_{n}+\psi_{n+L^2}^b(i)B_{n}]
\eea
and the diagonalized form is
\be
H_{diag} = \sum_n \lbs E_n^a A^\dagger_n A_n +E_n^b B^\dagger_n B_n\rbs
\ee
Here $E_n^a$'s are positive eigenvalues and $E_n^b$'s are negative eigenvalues.
In the transformed basis, the current operator $\frac{ i J_x(i)}{e t_0}$ can be written as \\
\bea
\sum_{j=1}^3\delta_{ij}^x \sum_{n,m}\lbs \psi_n^a(i)\psi_m^b(j)-\psi_n^b(j)\psi_m^a((
i)\rbs A^\dagger_n A_m \nonumber \\
+\sum_{j=1}^3\delta_{ij}^x \sum_{n,m} \lbs \psi_{n^\prime}^a(i)\psi_{m^\prime}^b(j)-\psi_{n^\prime}^b(j)\psi_{m^\prime}^a(i)\rbs B^\dagger_n B_m \nonumber \\
+\sum_{j=1}^3\delta_{ij}^x \sum_{n,m}\lbs \psi_n^a(i)\psi_{m^\prime}^b(j)-\psi_n^b(j)\psi_{m^\prime}^a(i)\rbs A^\dagger_nB_m \nonumber \\
+\sum_{j=1}^3\delta_{ij}^x \sum_{n,m} \lbs \psi_{n^\prime}^a(i)\psi_m^b(j)-\psi_{n^\prime}^b(j)\psi_m^a(i)\rbs B^\dagger_n A_m \nonumber \\
\eea
Here $n^\prime=n+L^2$ and $m^\prime=m+L^2$. 
Using this expression for the current operator and with some simple algebra, we get the following Lehmann representation for the current-current correlation function
\be
\Lambda_{xx}(\omega) = \frac{e^2}{2L^2}\sum\limits_{n=1,m=1}^{L^2}\sum_{p=1}^4 F_p(n,m)
\ee
where $F_1(n,m)$ is given by
\be
\frac{f_n^a-f_m^a}{\omega+E_n^a-E_m^a+i \eta}[\sum\limits_{<i,j>}\delta_j^x(\psi_n^a(i)\psi_m^b(j)-\psi_m^a(i)\psi_n^b(j))]^2
\ee
Similarly $F_4(n,m)$ is given by
\be
\frac{f_{n}^b-f_{m}^a}{\omega+E_{n}^b-E_{m}^a+i \eta}[\sum\limits_{<i,j>}\delta_j^x(\psi_{n^\prime}^a(i)\psi_{m}^b(j)-\psi_{m}^a(i)\psi_{n^\prime}^b(j))]^2]
\ee
$F_2(n,m)$ can be obtained from $F_1(n,m)$ by replacing $n, m$ in the indices for wave functions by $n^\prime, m^\prime$ which also implies replacing $E^a_n$ by $E^b_n$ and $f_n^a$ by $f_n^b$ respectively. Similarly $F_3(n,m)$ can be obtained from $F_4(n,m)$ by replacing $n^\prime$ by $n$ and $m$ by $m^\prime$. Correspondingly $E^a$ and $f^a$ are replaced by $E^b$ and $f^b$ and vice versa. 
This is the expression after taking the $q\rightarrow 0$ limit. Here $f_n^{a,b} = 1/(\exp(\beta (E_n^{a,b}-\mu))+1)$ is the Fermi function and $\eta \rightarrow 0^{+}$. Finally we substitute for $\Lambda_{xx}$ and calculate the conductivity using Eq.[\ref{sigma}].


\begin{thebibliography}{99}
\bibitem{footnote}
Though $\xi_l$ from the time evolution dynamics provides qualitatively the same picture as we obtained from the IPR analysis, there are quantitative differences. The location of the mobility edge for a given disorder are slighly different in two calculations. Main reason for this might be less number of configuration averaging for the time evolution dynamics (as it is very time consuming) compared to the IPR analysis and different bin sizes used in two calculations.
\end{thebibliography}
\end{document}